\newif\ifproofs
\proofstrue 

\documentclass{article}
\usepackage{Shared/spconf}
\usepackage{Shared/NUV_DoA}
\usepackage{cite}
\usepackage[table, dvipsnames]{xcolor}
\usepackage[bookmarks,colorlinks]{hyperref}
\usepackage{subcaption}
\usepackage{bbm}

\title{NUV-DOA: NUV Prior-based Bayesian Sparse Reconstruction with Spatial Filtering for Super-Resolution DoA Estimation}
\name{Mengyuan Zhao, Guy Revach, Tirza Routtenberg, and Nir Shlezinger
\thanks{
M. Zhao and G. Revach contributed equally to this work. 
M. Zhao was doing this work while at ETH Zürich, now at the Division of ISE, KTH Royal Institute of Technology, Sweden. G. Revach is with ISI, D-ITET, ETH Zürich (email: grevach@ethz.ch), 
T. Routtenberg and N. Shlezinger are with the School of ECE, Ben-Gurion University of the Negev, Be`er Sheva, Israel.
We thank Hans-Andrea Loeliger for the helpful discussions.}}
\address{\vspace{-25mm}}
\begin{document}
%
\maketitle
%
%
\begin{abstract}
%
 Achieving high-resolution \ac{doa} recovery 
typically 
requires high \ac{snr} and a sufficiently large number of snapshots. This paper presents \acl{ndoa} algorithm, that augments Bayesian sparse reconstruction with spatial filtering for super-resolution \ac{doa} estimation. By modeling each direction on the azimuth's grid with the sparsity-promoting \ac{nuv} prior, the non-convex optimization problem is reduced to iteratively reweighted \acl{ls} under Gaussian distribution, where the mean of the snapshots is a sufficient statistic. This approach not only simplifies our solution but also accurately detects the \acp{doa}. We utilize a hierarchical approach for interference cancellation in multi-source scenarios. Empirical evaluations show the superiority of \acl{ndoa}, especially in low \acsp{snr}, compared to alternative \ac{doa} estimators.
\end{abstract}
%
%
\acresetall 
%
%
\vspace{-0.2cm}
\section{Introduction}\label{sec:intro}
\vspace{-0.15cm}
\ac{doa} estimation is the task of determining {the azimuth of source signals using multiple measurements from} a sensor array~\cite{benesty2017fundamentals}. It plays a crucial role in contemporary applications across various fields~\cite{ThreeDecades_23}.
While the resolution of conventional approaches  based on covariance recovery via beamforming~\cite{bartlett1948beamformer} and \acs{mvdr}~\cite{capon1969mvdrbf}, is limited by the array geometry, subspace methods (e.g., \acs{music} \cite{schmidt1986music}) achieve super-resolution. However, they rely on a relatively large number of snapshots and  high \acp{snr}, and their accuracy degrades when these  are not met.


To address some of the limitations of previous approaches, \acl{dl}-based methods have emerged. These methods utilize either \acl{e2e} \aclp{nn} for direct \ac{doa} recovery~\cite{doadnn2018liu, convnet2019chakrabarty, Elbir20, Papageorgiou2020DeepNF} or augment classic algorithms~\cite{damusic2022merkofer,  shmuel2023subspacenet}, with trainable architectures via \acl{mb} \acl{dl}~\cite{SIG-113}. While these methods often function effectively under challenging conditions, they are inherently \acl{dd},  relying on labeled data from which the estimation mapping is learned. 


The unique structure of the \ac{doa} estimation problem has facilitated the adoption of  \ac{ssr}~\cite{BP_2001} and \ac{cs}~\cite{CS_06_trans_inf}. In this approach, the azimuth range is quantized into a grid, with the steering vectors serving as the dictionary. Methods such as $\ell_1$-SVD~\cite{L1SVD_05} and SPICE~\cite{SPCICE_11}, as well as others~\cite{CS_MUSIC_11, shen2016underdetermined}, have demonstrated their merit in challenging scenarios, particularly those with few snapshots or correlated sources~\cite{ThreeDecades_23}.
%
%
Within the scope of \textit{sparse} modeling, the \textit{Bayesian} framework~\cite{bayesianinterpolation1992mackay} can be interpreted as a form of regularization where sparsity is promoted through the adoption of specific prior distributions. Notably, methods such as \acs{rvm}, \acs{sbl}, and \acs{bcs}~\cite{RVM_2001, SBL_2004, BCS_2008}  present an effective surrogate for addressing $\ell_0$ minimization. This often results in more accurate recovery performance compared to many $\ell_1$ minimization approaches. Consequently,  Bayesian-based \ac{doa} estimation techniques have been proposed, including~\cite{wipf2007beamforming, liu2012sparsebayes, Yang2013offgridDoA, bayescompsens2013carlin, 7536146, ZHANG2016153}. Although these methods all stem from a shared concept, they exhibit distinct differences in modeling, choice of sparsifying prior, estimation objective, choice of estimation algorithm, tuning parameters, the handling of \ac{mmv}, and decision rules. These nuances  impact resolution, recovery performance, stability, and  complexity.
%
%

Despite their potential, \ac{ssr}-based methods exhibit inherent limitations arising from the unique structure of the \ac{doa} estimation problem and the associated resolution trade-off. Achieving higher resolution necessitates refining the grid quantization. However, this refinement leads to a more complex problem and a dictionary with greater mutual coherence, which can, in turn, impair the recovery performance. A further limitation arises from the handling of \ac{mmv}. While some approaches address \ac{mmv} by combining multiple single-snapshot (\acs{smv}) estimators, others tackle the full \ac{mmv} problem, frequently incorporating joint sparsity. However, these methods often increase the problem's complexity proportional to the number of snapshots, rendering them less efficient.
%
%

To tackle these challenges, we present \textit{\acl{ndoa}}, a super-resolution \ac{ssr}-based \ac{doa} estimation algorithm that utilizes the mean of the snapshots as its sufficient statistic. Despite its simplicity, it offers high accuracy and robustness, especially in scenarios with low \ac{snr} and a limited number of snapshots. In this approach, each quantized cell in the azimuth grid is modeled as a complex decision variable, adhering to a sparsity-inducing \ac{nuv} prior~\cite{Loeliger2017, Loeliger2019, Loeliger2023}. The \ac{nuv}, deeply rooted in \textit{Bayesian} sparsity, facilitates the transformation of the \ac{ssr} problem into an iteratively reweighted \acl{ls} Gaussian estimation problem, where the mean of the snapshots serves as a sufficient statistic. To further improve the \ac{doa} recovery performance, we harness the inherent spatial correlation of the \ac{doa} estimation problem. We employ a spatial filtering-based super-resolution algorithm that segments a single, global \ac{ssr} problem spanning the entire grid into a series of localized problems. Finally, we propose a hierarchical algorithm designed to reduce computational complexity for super-resolution. This approach is especially vital in scenarios with multiple sources. By incorporating an interference cancellation step, we simplify the multi-source \ac{doa} recovery into 
$K$ single-source problems without incurring additional complexity.

The remainder of this paper is structured as follows: Section \ref{sec: pre} formulates the problem of \ac{doa} estimation and outlines its sparse representation; Sections \ref{sec: NUV-EM alg}-\ref{sec:emp_eval} describe and evaluate \acl{ndoa}, respectively, while Section~\ref{sec:Conclusions} concludes the paper. 
\vspace{-0.2cm}
\section{Problem Formulation and Model}\label{sec: pre}
\vspace{-0.15cm}
%
%
\subsection{DoA Estimation Problem Formulation}
\label{ssec:problem formulation}
\vspace{-0.1cm}
We study the  estimation of the \acp{doa} of $K$ far-field radio signals originating from directions $\boldsymbol{\theta}\in\Theta^K$, where $\Theta= \left[-\frac{\pi}{2}, \frac{\pi}{2}\right)$. The \acp{doa} are recovered from $L$ noisy snapshots measured using a \ac{ula} consisting of $N$ elements spaced at half-wavelength intervals. A single snapshot $\gvec{y}(t)$ can be expressed as
\vspace{-0.1cm}
\begin{equation}
\gvec{y}(t) = \gvec{A}\brackets{\boldsymbol{\theta}}\cdot \gvec{s}\brackets{t}
+ \gvec{v}\brackets{t}
\in\gcmplx^N,
\quad
t\in\set{1,...,L}.
\label{doa generate one}
\vspace{-0.1cm}
\end{equation}
Here, $\gvec{A}\brackets{\boldsymbol{\theta}}\in\gcmplx^{N\times K}$ represents the steering matrix, which projects the $K$ narrow-band impinging source signals, denoted by vector $\gvec{s}\brackets{t}\in\gcmplx^K$, onto the \ac{ula}. 
The steering matrix is comprised of the $K$ steering vectors, defined as
\vspace{-0.1cm}
\begin{equation}\label{eq:steer_vec}
\gvec{a}\brackets{\theta_k}= 
\brackets{1, e^{-i\pi\sin\brackets{\theta_k}},\ldots,
e^{-i\pi\brackets{N-1}\sin\brackets{\theta_k}}}^\top.
\vspace{-0.1cm}
\end{equation}
The term $\gvec{v}\brackets{t}$ in \eqref{doa generate one} is an additive  noise following a complex Gaussian distribution, $\gcnormal{0,\gvec{R}}$.
%
%
%
Aggregating the $L$ snapshots as the matrix  $\gvec{Y}$, our objective is to estimate  $\boldsymbol{\theta}$ from $\gvec{Y}$. 

%
%
\vspace{-0.1cm}
\subsection{Sparse Modeling}
\label{ssec: sparse doa}
\vspace{-0.1cm}
To frame the \ac{doa} estimation problem as an instance of \ac{ssr}, we quantize the interval $\Theta$ into $M$ equidistant grid cells
\vspace{-0.1cm}
\begin{equation*}
\boldsymbol{\vartheta} \triangleq \brackets{\vartheta_0,\ldots,\vartheta_{M-1}},
\hspace{0.1cm}
\vartheta_m=m\cdot\Delta\vartheta-\frac{\pi}{2},
\hspace{0.1cm}
\Delta\vartheta=\frac{\pi}{M}.
\vspace{-0.1cm}
\end{equation*}
Using these quantized directions, we can approximate a snapshot as a \textit{sparse} linear combination of steering vectors. Specifically, let us denote
\vspace{-0.1cm}
\begin{equation}
\label{eqn:SSRmodel}
\hat{\gvec{y}}\brackets{t}=
\strmat\cdot\gvec{x}\brackets{t},\hspace{0.2cm}
\strmat\in\gcmplx^{N\times M},
\hspace{0.2cm}
\gvec{x}\brackets{t}\in\gcmplx^M,
\vspace{-0.1cm}
\end{equation}
where $\strmat$ is an over-complete dictionary matrix with the $m\textrm{-th}$ column being a steering vector represented by $\gvec{a}\brackets{\vartheta_m}$ as in \eqref{eq:steer_vec}. Furthermore, $\gvec{x}\brackets{t}$ is a sparse vector in which each non-zero entry corresponds to an active direction. 
The \acs{smv} \ac{ssr} optimization problem  with a single realization $\gvec{y}$ (omitting $t$ for brevity), is represented as:
\vspace{-0.1cm}
\begin{equation}
\gvec{x}^\ast=\arg\min_{\gvec{x}}
\set{\norm{\gvec{y}-\strmat\cdot\gvec{x}}_2^2+ \gamma\norm{\gvec{x}}_0}.
\label{eq:SMV_SSR}
\vspace{-0.1cm}
\end{equation}
Problem~\eqref{eq:SMV_SSR} is  challenging to address due to the non-convexity introduced by the $\ell_0$ norm. Several approaches have been proposed in the literature for solving this problem , including greedy OMP methods~\cite{tropp2007signal}, optimization-based methods using $\ell_1$ and $\ell_2$ relaxations \cite{BP_2001}, and \textit{Bayesian} methods~\cite{SBL_2004}. For \ac{mmv}~\cite{MMV_05}, the conventional method employs a block of measurements coupled with a block of sparse vectors, assuming joint sparsity. Yet, this method sees its complexity increase with the number of snapshots. In the following, we demonstrate that by adopting a \textit{Bayesian} approach with the \ac{nuv} prior, the \ac{mmv} problem can be distilled down to an \acs{smv}. In this context, $\bar{\gvec{y}}_L$, representing the mean of multiple snapshots, emerges as a sufficient statistic.
%
%
%
%
%
\begin{figure*}[!t]
\scriptsize
\centering
\begin{minipage}{.125\linewidth}
\centering
\includegraphics[width=1\columnwidth]{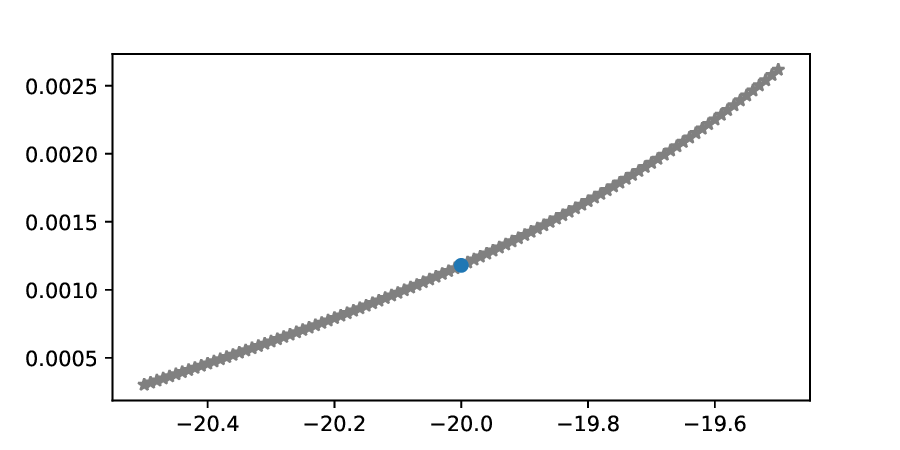}
\captionsetup{font=footnotesize}
\caption{$\theta=-20^\circ$}
\label{fig:SCL}
\end{minipage}
\begin{minipage}{.125\linewidth}
\centering
\includegraphics[width=1\columnwidth]
{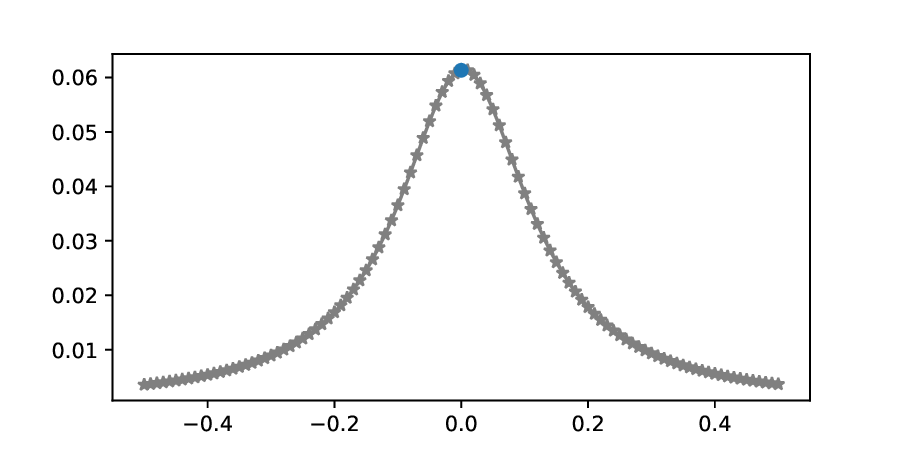}
\captionsetup{font=footnotesize}
\caption{$\theta=0^\circ$}
\label{fig:SCC}
\end{minipage}
\begin{minipage}{.125\linewidth}
\centering
\includegraphics[width=1\columnwidth]
{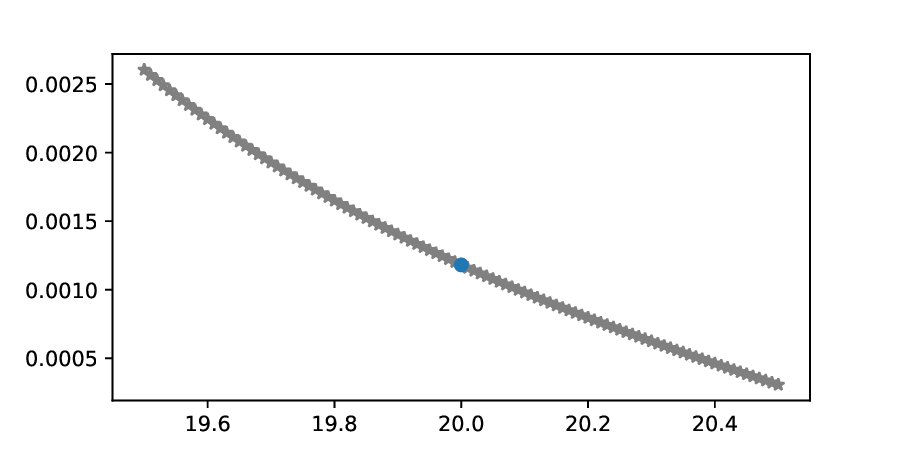}
\captionsetup{font=footnotesize}
\caption{$\theta=+20^\circ$}
\label{fig:SCR}
\end{minipage}
\begin{minipage}{.23\linewidth}
\centering
\includegraphics[width=1\columnwidth]
{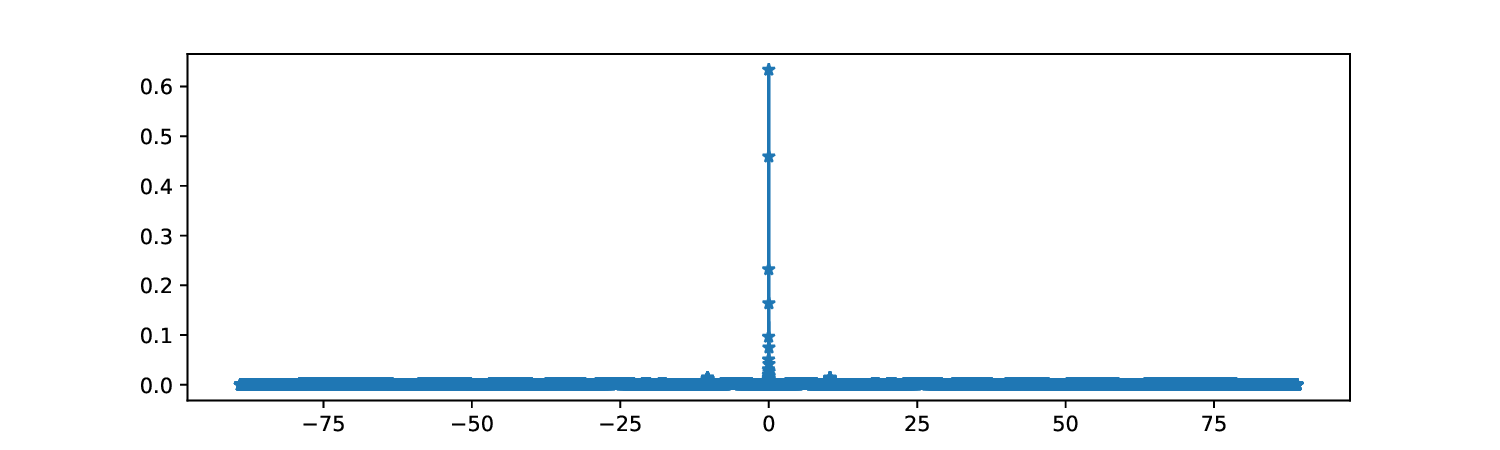}
\captionsetup{font=footnotesize}
\caption{$\theta\leq\abs{75}^\circ$}
\label{fig:SCT}
\end{minipage}
\begin{minipage}{.125\linewidth}
\centering
\includegraphics[width=1\columnwidth]
{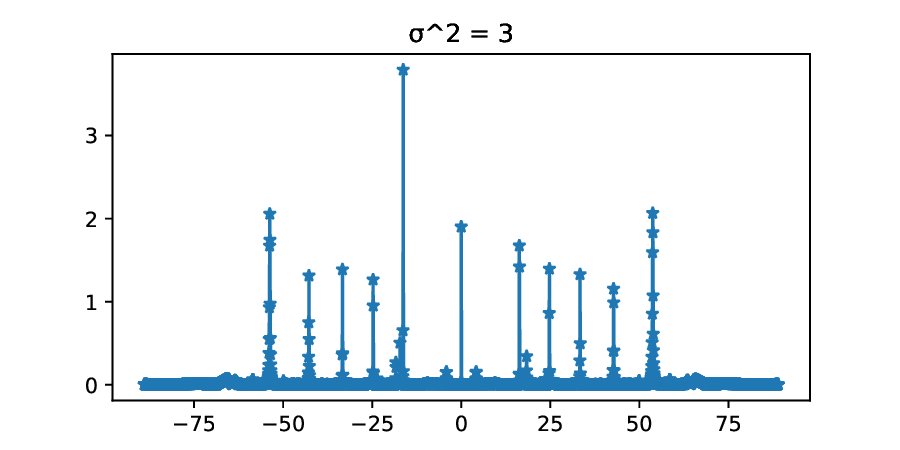}
\captionsetup{font=footnotesize}
\caption{$\sigma^2=3\cdot10^0$}
\label{fig:tun_L}
\end{minipage}
\begin{minipage}{.125\linewidth}
\centering
\includegraphics[width=1\columnwidth]
{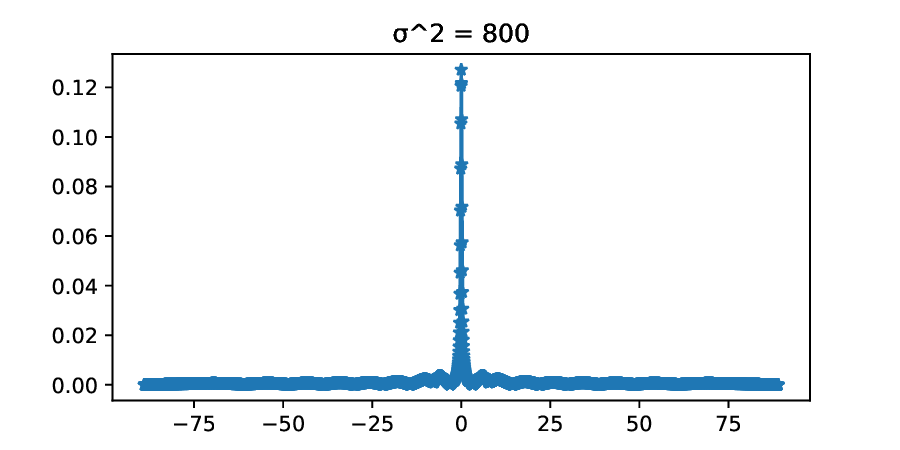}
\captionsetup{font=footnotesize}
\caption{$\sigma^2=8\cdot10^2$}
\label{fig:tun_M}
\end{minipage}
\begin{minipage}{.125\linewidth}
\centering
\includegraphics[width=1\columnwidth]
{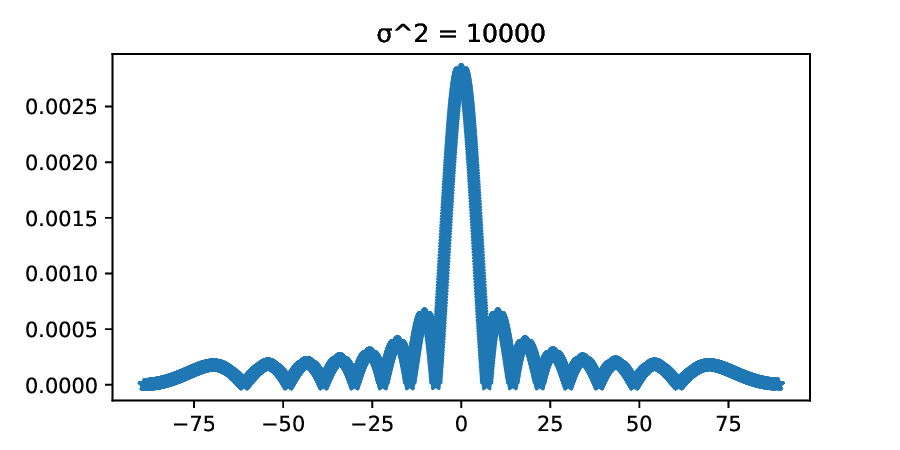}
\captionsetup{font=footnotesize}
\caption{$\sigma^2=10^4$}
\label{fig:tun_H}
\end{minipage}
\end{figure*}
%
%
\vspace{-0.2cm}
\section{NUV-DoA High-Resolution Estimation}
\label{sec: NUV-EM alg}
\vspace{-0.15cm}
Our \acl{ndoa} algorithm spans three key ingredients: 1) the \acl{nssr}, a general Bayesian-based \ac{ssr} algorithm leveraging the \ac{nuv} prior to promote sparsity that is versatile enough to be applied beyond \ac{doa} estimation to any dictionary matrix; 2) a \textit{super-resolution} algorithm that capitalizes on the inherent spatial correlation of the DoA setting through spatial filtering over overlapping sub-bands; and 3) a hierarchical strategy for multi-source \textit{interference cancellation}.
%
%
\vspace{-0.1cm}
\subsection{NUV-SSR}
\vspace{-0.1cm}
In our model \eqref{eqn:SSRmodel}, it holds that the vectors $\set{\gvec{x}\brackets{t}}_{t=1}^L$ are jointly sparse; that is, they all share the same support. Moreover, every snapshot is considered independent and of equal importance. Consequently, we represent $\gvec{x}\in\mathbb C^M$ as a vector of complex decision variables, where each component $\gscal{x}_m\in\gcmplx$ corresponds to a potential hypothesis, such as a direction $\vartheta_m$, and follows a complex \ac{nuv} prior, specifically: 
\vspace{-0.1cm}
\begin{equation}
\gscal{x}_m\sim\gcnormal{0, \gscal{q}_m^2},
\quad
\prob{\gscal{q}_m^2}\propto 1, 
\quad
\gscal{q}^2_m\in\greal_+.
\end{equation}
\vspace{-0.1cm}
The joint likelihood of the snapshots, given the decision vector, can be decomposed into conditionally independent and identical Gaussian distributions. Given that our prior adheres to a Gaussian distribution, the posterior is Gaussian as well. Thus, the temporal mean  $\bar{\gvec{y}}_L = \frac{1}{L}\sum_{t=1}^L\gvec{y}(t)$, emerges as a sufficient statistic, significantly simplifying  inference.

The conventional approach for recovery involves a Type I method: specifically, a \ac{map} estimation of $\gvec{x}$ given the sufficient statistic $\bar{\gvec{y}}_L$. However, studies have shown that Type II offers superior recovery performance~\cite{TypeII_2016}. In Type I, the vector of unknown variances $\gvec{q}^2=\brackets{\gscal{q}^2_1,\ldots,\gscal{q}^2_M}^\top$ is integrated out, while here, in Type II, it is estimated through evident maximization: 
\begin{equation}
\hat{\gvec{q}}^2=\arg\max_{\gvec{q}^2\in\greal_+^M}
\set{\prob{\gvec{q}^2\given{\bar{\gvec{y}}_L}}}.
\label{eq:MAP} 
\end{equation}
To solve this optimization problem, we employ a two-step \ac{em} algorithm~\cite{shumway1982approach}. By considering $\gvec{x}$ as a latent vector dependent on $\gvec{q}^2$ and maximizing the log-posterior lower bound difference, the algorithm reduces to an iteratively reweighted \acl{ls} estimation of the unknown variance, based on the following update rule:
%
%
\begin{equation}
\hat{\gscal{q}}^2_{m,\sbrackets{i+1}}
=\expval_{\sbrackets{i}}\brackets{\abs{\gscal{x}_m}^2}=
\abs{\expval_{\sbrackets{i}}\brackets{\gscal{x}_m}}^2+
\varval_{\sbrackets{i}}\brackets{\gscal{x}_m}.
\end{equation}
%
Here, $\expval_{\sbrackets{i}}$, and $\varval_{\sbrackets{i}}$ are the expectation and variance, respectively, computed under the posterior distribution $\prob{\gvec{x}\given{\bar{\gvec{y}}_L;\hat{\gvec{q}}^2_{\sbrackets{i}}}}$, with the \acp{nuv} set to their values from the preceding iteration. These values arise from minimizing a \acl{ls} cost function, for which the closed-form solution is
\begin{subequations}
\begin{align}
&\hat{\gvec{x}}_{\sbrackets{i}}=
\expval_{\sbrackets{i}}\brackets{\gvec{x}}=
\Gamma\brackets{\hat{\gvec{q}}^2_{\sbrackets{i}}}\cdot \gvec{A}^\herm\cdot 
\gvec{W}_{\sbrackets{i}} \cdot \bar{\gvec{y}}_L,\\
&\varval_{\sbrackets{i}}\brackets{\gvec{x}}=
\hat{\gvec{q}}^2_{\sbrackets{i}}-
\Gamma\brackets{\hat{\gvec{q}}^4_{\sbrackets{i}}}\cdot
\diag\brackets{\gvec{A}^\herm\cdot \gvec{W}_{\sbrackets{i}} \cdot\gvec{A}},
\end{align} 
\end{subequations}         
where $\gvec{W}_{\sbrackets{i}}$ is a precision matrix of size $n\times n$ given by:
\begin{equation}
\gvec{W}_{\sbrackets{i}} = 
\brackets{\gvec{A}\cdot
\Gamma\brackets{\hat{\gvec{q}}^2_{\sbrackets{i}}}
\cdot
\gvec{A}^\herm
+
\frac{\sigma^2}{L}\cdot\mathbf{I}}^{-1}.
\end{equation}
In the above, $\brackets{\boldsymbol{\vartheta}}$ was omitted to indicate that $\gvec{A}$ can serve as a general
dictionary. $\Gamma\brackets{\gvec{u}}$ is  a diagonal matrix with $\gvec{u}$  being its main diagonal, and $\diag\brackets{\gvec{B}}$ denotes an operator of extracting the diagonal of matrix $\gvec{B}$ and formatting it as a vector.  The \acl{nssr} is initialized with $\hat{\gvec{q}}^2_{\sbrackets{0}}$ as a non-zero random value, and continues until a predefined convergence criterion is met. A notable benefit of this method lies in its single-sparsity tuning parameter 
$\sigma^2$, further elaborated upon in \secref{sec:emp_eval}.
%
%

A key feature of the \ac{nuv} representation is that 
when $\gscal{q}^2_m=0$, it implies that $\abs{\gscal{x}_m}=0$ as well. To select the $K$ active hypotheses, we introduce $\gvec{\Omega}\in\mathbb R^M$ as the \acl{nus} with $\Omega_m=\abs{\gscal{x}_m}$, $m=1,...,M$, from which we choose the $K$ most dominant peaks. If $K$ is not pre-defined, we select the most dominant peaks that exceed a threshold $\eta$.
%
%
%
\begin{figure*}[!t]
\scriptsize
\centering
\begin{minipage}{.245\linewidth}
\centering
\includegraphics[width=1\columnwidth]{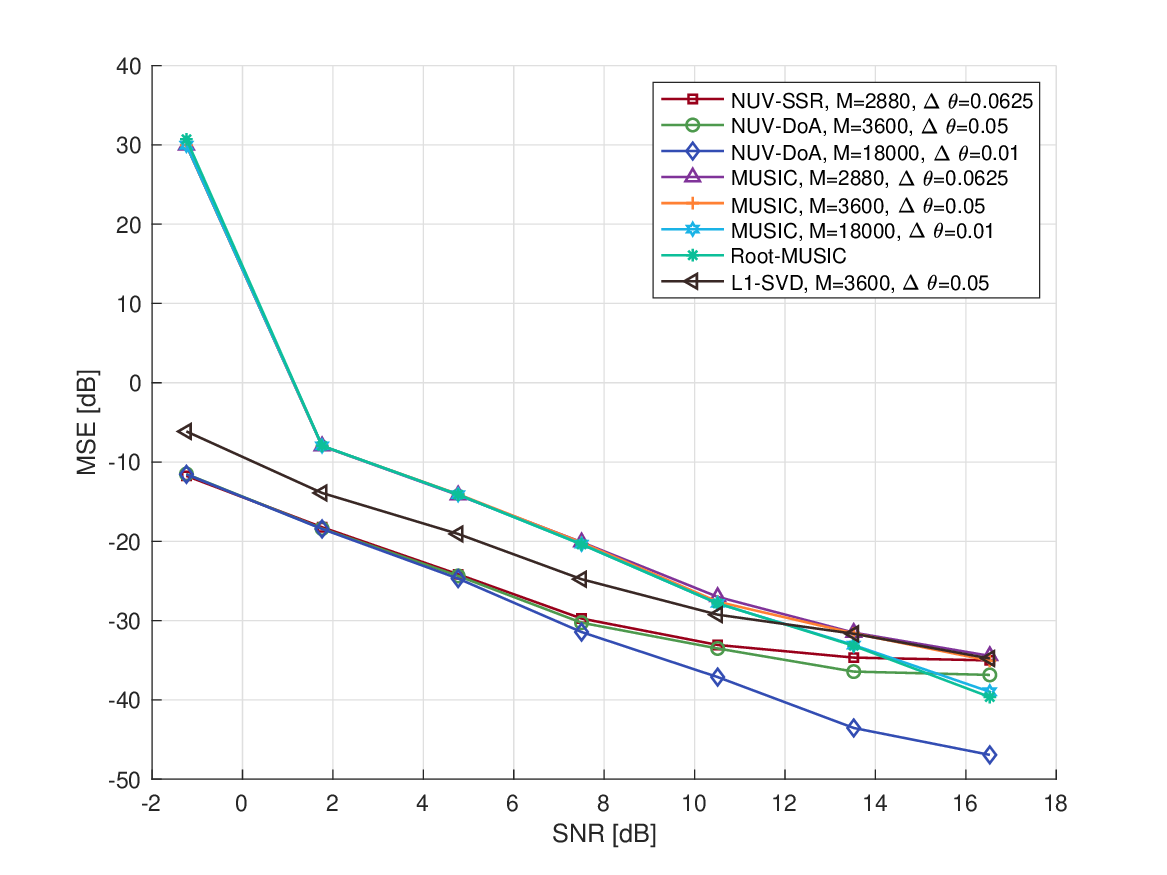}
\captionsetup{font=footnotesize}
\caption{$L=100$, $\theta\leq\abs{75}^\circ$}
\label{fig:L_100_center}
\end{minipage}
\begin{minipage}{.245\linewidth}
\centering
\includegraphics[width=1\columnwidth]{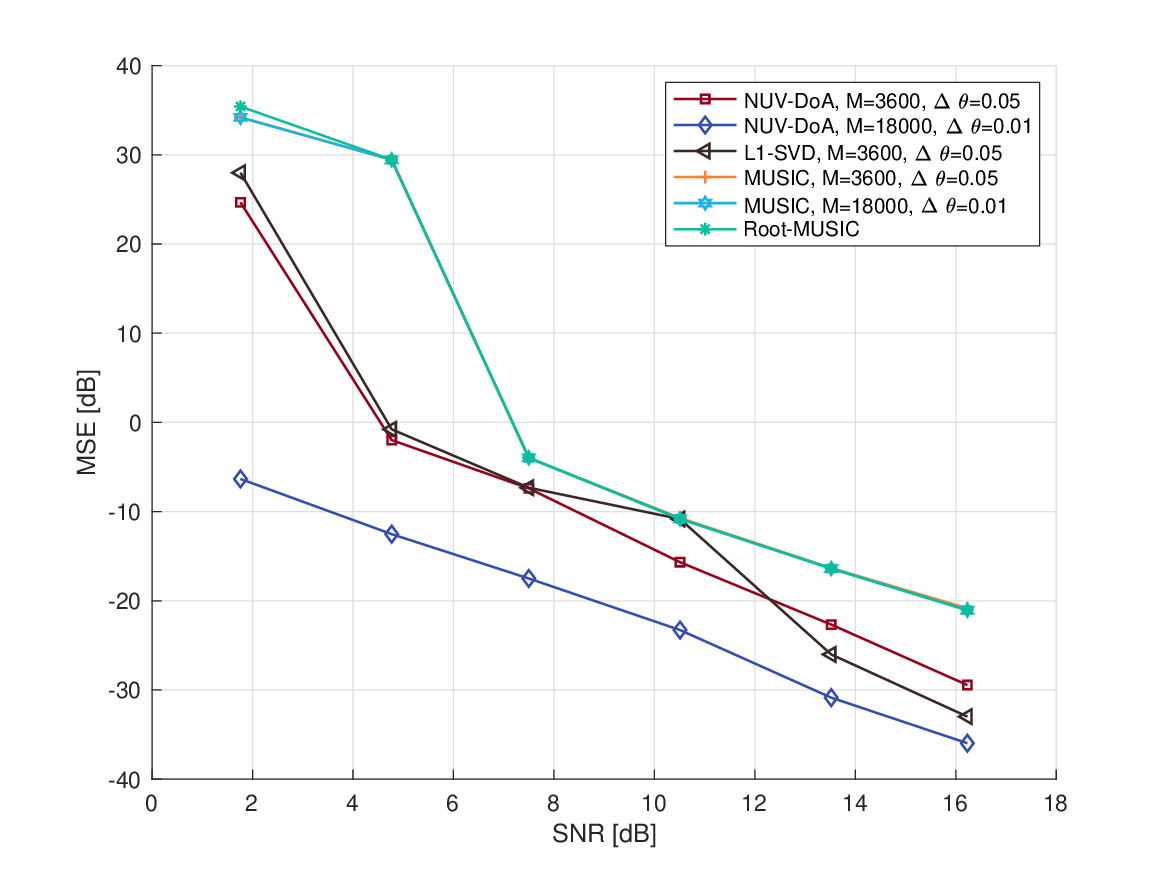}
\captionsetup{font=footnotesize}
\caption{$L=100$, $\abs{75}^\circ\leq\theta\leq\abs{85}^\circ$}
\label{fig:L_100_bounds}
\end{minipage}
\begin{minipage}{.245\linewidth}
\centering
\includegraphics[width=1\columnwidth]{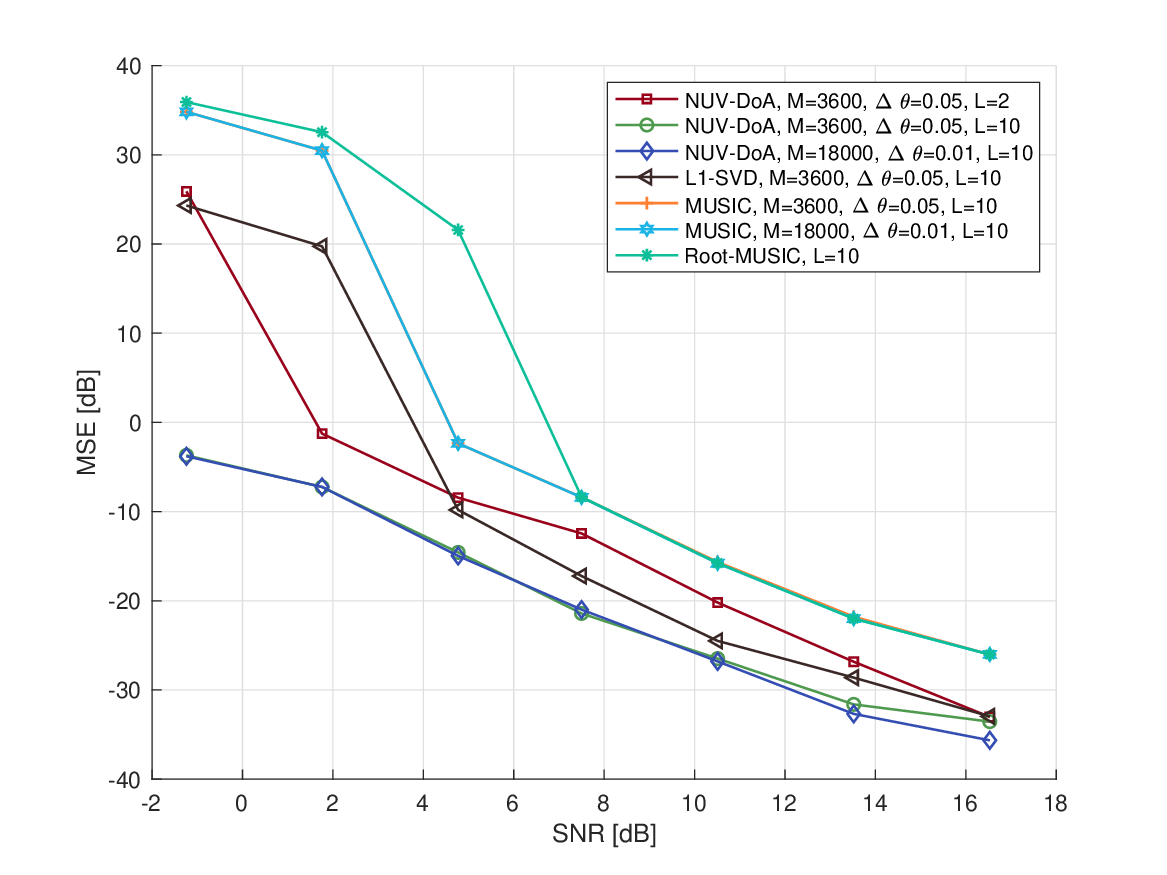}
\captionsetup{font=footnotesize}
\caption{$L=2,10$, $\theta\leq\abs{75}^\circ$}
\label{fig:L10}
\end{minipage}
\begin{minipage}{.245\linewidth}
\centering
\includegraphics[width=1\columnwidth]{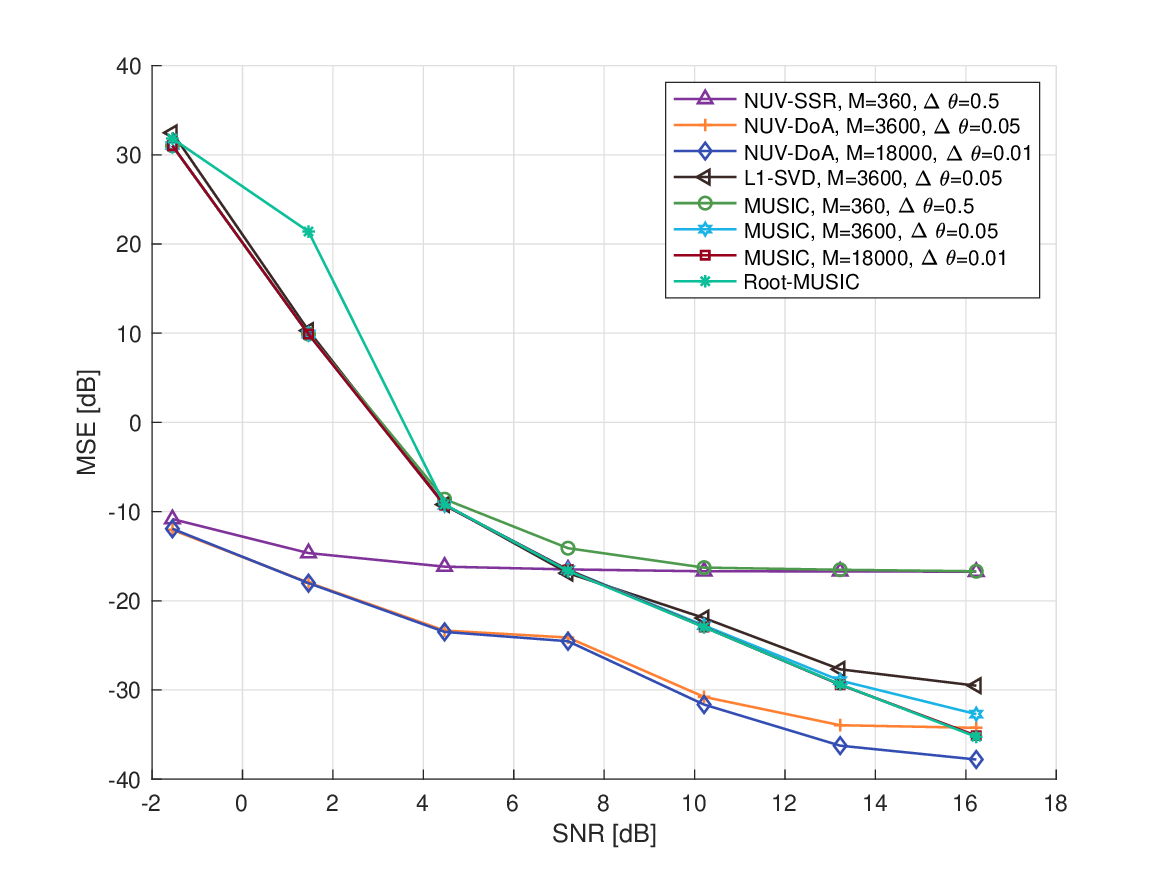}
\captionsetup{font=footnotesize}
\caption{$K=2$, $L=100$, $\theta\leq\abs{75}^\circ$}
\label{fig:K2}
\end{minipage}
\end{figure*}
%
%
%
\vspace{-0.1cm}
\subsection{Super-Resolution by Spatial Filtering}
\label{ssec:spec_combine}
\vspace{-0.1cm}
The outlined \acl{nssr} is  powerful and robust, and we  successfully employed it to address an \ac{ssr} problem encompassing $M=3000$ atoms. Despite the computational complexity of the problem, the performance is commendable. In  \ac{doa} recovery, $M=3000$ translates to a resolution of $\Delta\vartheta=0.06^\circ$. However, aiming for finer resolutions like $\Delta\vartheta=0.01^\circ$ dramatically inflates the problem size to $M=18,000$ atoms, presenting substantial challenges to all \ac{ssr} algorithms. The unique structure of the \ac{doa} recovery problem further complicates matters: the steering vectors exhibit spatial correlation, thus increasing the dictionary's mutual coherence which can potentially reduce performance.
%
%

To further enhance \ac{doa} recovery performance, we augment \acl{nssr} with super-resolution by leveraging the inherent spatial correlation, leading to the implementation of \acl{ndoa} by spatial filtering. Here, we segment the single, global \ac{ssr} problem spanning the entire azimuth grid into a series of smaller localized overlapping tasks. Each task targets a sub-band of the azimuth around $\vartheta_m$ of size $2\alpha$, represented as
$\vartheta^{\sqcap}_m=\sbrackets{\vartheta_m-\alpha,\vartheta_m+\alpha}$. For every sub-band $\vartheta^{\sqcap}_m$, an \ac{ssr} problem is solved based on the steering sub-matrix $\gvec{A}^\sqcap_m\brackets{\boldsymbol{\vartheta}}$ using \acl{nssr}. Within each of these sub-bands, only the center element from the sub-spectrum $\gvec{\Omega}^\sqcap_m$ is extracted. Following this, these elements are combined to produce the full spectrum that spans the entire azimuth, after which peak detection is executed.
%
%

This strategy excels in delivering high resolution, precision, and a universal tuning principle. When targeting a resolution of $\Delta\vartheta=0.01^\circ$ with $\alpha=0.5^\circ$, each sub-band spans $1^\circ$. This results in $M^\sqcap=101$ decision variables for each sub-task. This method proves more effective and efficient than other alternatives. Yet, its efficiency can be further enhanced by adopting a hierarchical approach, as we show next.
%
%
%
%
\vspace{-0.1cm}
\subsection{Hierarchical Approach and Multi-Source Estimation}
\vspace{-0.1cm}
While the spatial filtering strategy significantly enhances resolution, it can also introduce an off-grid effect. Energy from sources located outside the band might leak into the sub-band of interest, causing interference. To address this, we employ a hierarchical approach for adjacent channel interference cancellation, transforming multi-source \ac{doa} recovery into $K$ single-source problems, without adding extra complexity.

We start with a low-resolution preprocessing phase to coarsely estimate the $K$ directions using \acl{nssr} in low \acp{snr}, and Root-MUSIC in high \acp{snr} (e.g., SNR threshold 7dB). For each single estimated source, interference from its $K-1$ neighboring sources is canceled by subtracting a linear combination of their steering vectors from the observations. This leaves a single source \ac{doa} estimation, to which our high-resolution algorithm can be applied. Given our reliable direction estimates, high-resolution processing is confined to a narrow azimuth sub-band of size $6\cdot\epsilon$ around the low-resolution estimate, where $\epsilon$ represents the empirical standard deviation of the low-resolution algorithm's error at the given \ac{snr}. This hierarchical method is also adaptable for single-source estimation, offering a significant reduction in complexity.
%
%
\vspace{-0.1cm}
\subsection{Discussion}
\vspace{-0.1cm}
While the sparsifying \ac{nuv} prior is deeply rooted in \acs{sbl}, its potential in the Gaussian setting has not been fully harnessed for general \ac{ssr} and \ac{doa} estimation tasks. The fact that the mean of the snapshots is  a sufficient statistic not only simplifies the problem but also enhances the robustness of our \acl{ndoa}, particularly in low \acp{snr} and with a limited number of snapshots. Additionally, since our method does not rely on the covariance of the snapshots, it can effectively manage scenarios with coherent sources. Having only a single tuning parameter $\sigma^2$, makes \acl{ndoa} easy to implement.

Here, estimating a positive variance directly correlates with the recovery of an atom from the support and with detecting a source signal in a particular direction. From a statistical perspective, our parameter estimation algorithm translates into a simultaneous statistical test spanning a family of $M$ latent hypotheses, of which only $K$ are active. Owing to the sparsity-enforcing nature of the \ac{nuv}, we attain high detection rates while maintaining low false alarms. 

\vspace{-0.2cm}
\section{Empirical Evaluation}\label{sec:emp_eval}
\vspace{-0.15cm}
%
%
Here, we empirically evaluate the properties and performance of our proposed \acl{ndoa}.\footnote{{The source code can be found at} \url{https://github.com/MengyuanZha0/ICASSP24-NUV-DoA}}

Next, we examine the impact of the tuning parameter $\sigma^2$. A low $\sigma^2$ means the \ac{nuv} does not promote sparsity, leading to an imprecise spectrum with numerous false alarms (\figref{fig:tun_L}). Conversely, a high $\sigma^2$ 
aggressively enforces sparsity, yielding a spectrum that is accurate but wide and of low magnitude (\figref{fig:tun_H}). An optimal $\sigma^2$ produces a spectrum that is precise, narrow, and of high magnitude, as seen in~\figref{fig:tun_M}.

We conclude by comparing the performance of \acl{ndoa}, across multiple \ac{snr} conditions, with benchmark algorithms. 
\figref{fig:L_100_center} reveals that our method, with a resolution of $\Delta\theta=0.01^\circ$, significantly surpasses its counterparts in both high and low \acp{snr}, especially when focusing on the mid-interval of the azimuth $\brackets{\theta\leq\abs{75}^\circ}$. This performance differential is even more pronounced at the azimuth boundaries $\brackets{\abs{75}^\circ\leq\theta\leq\abs{85}^\circ}$ as seen in~\figref{fig:L_100_bounds}. Notably, while competing algorithms falter even in medium \ac{snr} or with limited snapshots (e.g., $L = 10$), \acl{ndoa} retains efficacy even at an extreme low of $L = 2$, as depicted in~\figref{fig:L10}. Lastly, \figref{fig:K2} highlights the efficacy of our \acl{ic} mechanism, particularly when contending with two non-coherent sources spaced as closely as $15^\circ$ apart.

\vspace{-0.2cm}
\section{Conclusions}\label{sec:Conclusions}
\vspace{-0.15cm}
We proposed \acs{ndoa}, a high precision \ac{doa} estimation algorithm.  \acs{ndoa} combines Bayesian sparse recovery using \ac{nuv} priors, alongside spatial filtering for super-resolution and hierarchical multi-source estimation. Our numerical results show the superiority of \acs{ndoa} in robustness and accuracy, particularly in low \acp{snr}.

\vspace{-0.2cm}
%
\bibliographystyle{IEEEtran}
\bibliography{IEEEabrv,main}
\ifproofs

%

%

\fi

\end{document}